\documentclass[]{aa}
\usepackage{graphicx}
\newcommand{\be}{\begin{equation}}
\newcommand{\ee}{\end{equation}}

\begin{document}

\title{Precessing microblazars and unidentified gamma-ray sources}

\author{M. M. Kaufman Bernad\'o\inst{1}, G. E. Romero\inst{1,}\thanks{Member of CONICET},
I.F. Mirabel\inst{2,3,}$^{\star}$}

\offprints{M. M. Kaufman Bernad\'o}

\institute{Instituto Argentino de Radioastronom\'{\i}a, C.C.5,
(1894) Villa Elisa, Buenos Aires, Argentina \and
CEA/DMS/DAPNIA/Service D'Astrophysique, Centre d'Etudes de Saclay,
F-91191 Gif-sur-Yvette, France \and Instituto de Astronom\'{\i}a y
F\'{\i}sica del Espacio/CONICET, C.C. 67, Suc. 28, Buenos Aires,
Argentina}

\date{Received / Accepted}
\abstract{The recent discovery by Paredes et al. (2000) of a
persistent microquasar that is positionally coincident with an
unidentified gamma-ray source has open the possibility that other
sources in the Third EGRET Catalog could be interpreted as
microquasars as well. In this letter we show that some variable
unidentified EGRET sources in the galactic plane could be produced
by faint, otherwise undetected microquasars with precessing jets.
When the jet points towards the observer, gamma-ray emission
resulting from upscattered stellar photons could be detectable
yielding a variable source with weak or undetectable counterpart
at longer wavelengths. Strategies for detecting
these``microblazars'' with forthcoming satellites are briefly
discussed. \keywords{X-ray binaries -- stars: early-type --
gamma-rays: observations -- gamma-rays: theory}}

\titlerunning{Microblazars and gamma-ray sources}
\authorrunning{M. M. Kaufman Bernad\'o et al.}

\maketitle

\section{Introduction}

The third and final EGRET catalog (Hartman et al. 1999) lists
about 170 gamma-ray sources not yet clearly identified with
objects detected at lower energies. Many of these sources
concentrate towards the galactic plane and correlate with the
spiral arms of the Galaxy, indicating a significant contribution
from Population I objects (Romero et al. 1999, Romero 2001). From
a purely statistical point of view, at least two different
populations can be identified: one of them possibly associated
with the Gould belt, a nearby star-forming region at $\sim600$ pc,
and the other formed by brighter sources found at lower latitudes
(Gehrels et al. 2000, Grenier 2000).

Several types of objects have been proposed as possible
counterparts of galactic gamma-ray sources, including early-type
stars (both isolated and forming binary systems), accreting
neutron stars, radio-quiet pulsars, interacting supernova
remnants, and black hole candidates (see Romero 2001 and
references therein). Very recently, Paredes et al. (2000), Grenier
(2001) and Romero (2001) have suggested that microquasars can also
be responsible for some unidentified sources. In particular,
Paredes et al. (2000) have proposed that the microquasar LS 5039,
a massive X-ray binary with persistent non-thermal radio emission,
is physically associated with the gamma-ray source 3EG J1824-1514.
They suggested that the observed gamma-ray flux is the result of
external Compton scattering of UV photons from the high-mass
stellar companion.

The fact that there is a significant number of variable gamma-ray
sources at low galactic latitudes (e.g. Torres et al. 2001, Romero
et al. 2001) along with the steady properties of gamma-ray pulsars
and interacting supernova remnants, makes the idea of gamma-ray
emitting microquasars particularly attractive. Mirabel \&
Rodr\'{\i}guez (1999) proposed that microquasars with jets forming
a small angle with the line of sight --by analogy with the unified
model for AGNs-- could appear as {\em microblazars}, namely, as
sources with highly variable and enhanced nonthermal flux due to
Doppler boosting.


In this letter we shall present a model for galactic variable
gamma-ray sources based on the idea of precessing microblazars. In
our model, the jet precession will be induced by the gravitational
torque of the stellar companion, which is in a non-coplanar orbit,
on the accretion disk of the primary (usually assumed to be a
stellar black hole). Gamma-rays can be produced by external
Compton scattering of UV stellar photons of the massive companions
by relativistic leptons (electrons and/or positrons) far from the
base of the jet. We shall discuss cases where the particles are in
the form of a ``blob'' ejected from the central engine and where
the particles are continuously injected in the form of a
persistent jet. In the case of high-mass binaries, this mechanism
for generation of gamma-rays is more efficient than Compton
scattering upon disk photons or self-synchrotron jet emission
(Georganopoulos et al. 2001a) and, in combination with the
gravitational effects of the stellar companion, can render
suitable variable gamma-ray sources. We shall briefly discuss how
these sources can be identified in the light of forthcoming
satellite experiments like INTEGRAL. Although microblazars have
been previously considered in the literature (e.g. Aharonian \&
Atoyan 1998, Mirabel \& Rodr\'{\i}guez 1999, Georganopoulos et al.
2001a), this is the first detailed discussion in terms of
unidentified gamma-ray sources.

\section{Variable gamma-ray emission from microblazars}

Let us consider a high-mass binary where accretion onto the
compact object (a black hole) results in the production of twin,
relativistic $e^+e^-$-pair jets propagating in opposite
directions. A sudden variation in the injection rate can result in
the formation of a blob (for which we adopt spherical geometry)
which propagates down the jet with a bulk Lorentz factor $\Gamma$.
We shall follow the general treatment derived by Georganopoulos et
al. (2001b) for external Compton scattering in extragalactic
blazars, adapting it to microblazars. In the blob frame the
relativistic leptons are considered to have an isotropic power-law
density distribution
    $n'(\gamma\:')=(k/4\pi)\gamma\:'^{-p}P(\gamma_{1},\gamma_{2},\gamma\:')$,
where $\gamma\:'$ is the Lorentz factor of the leptons, \textit{k}
is a constant and $P(\gamma_{1},\gamma_{2},\gamma\:')=1$ for
$\gamma_{1}\leq\gamma\:'\leq\gamma_{2}$, and $0$ otherwise.

Using the Lorentz invariant $n/\gamma^{2}$ and the relation
$\gamma=D\gamma\:'$ between the Lorentz factor of the
electrons/positrons in the lab frame and the blob frame
respectively, we get that the electron density in the lab frame is
\begin{eqnarray}
    n(\gamma)=\frac{k}{4\pi}D^{2+p}\gamma^{-p}P(\gamma_{1}D,\gamma_{2}D,\gamma) \label{den},
\end{eqnarray}
where $ D=\left[\Gamma\left(1-\beta\cos\phi\right)\right]^{-1}$ is
the usual Doppler factor: $\beta$ and $\phi$ are the bulk velocity
in units of $c$ and the viewing angle, respectively.

The blob is then injected into an isotropic monoenergetic photon
field due to the companion star (typically UV photons) and will
then produce the inverse Compton upscattering of some of these
stellar photons. With respect to the photon field, the lab frame
rate of IC interactions per final photon energy is,
\begin{eqnarray}
    \frac{dN_{p}}{dtd{\epsilon}}=\frac{3\sigma_{T}c}{4\epsilon_{0}\gamma^{2}}f(x)
    \label{rate}
\end{eqnarray}
where $\sigma_{T}$ is the Thomson cross section. For the case of
Thomson scattering and assuming isotropic scattering in the
electron frame (Georganopoulos et al 2001b),
\begin{eqnarray}
    f(x)=\frac{2}{3}(1-x)P(1/4\gamma^{2},1,x),\ \ \
    x=\frac{\epsilon}{4\gamma^{2}\epsilon_{0}}.
\end{eqnarray}

\begin{figure}
\resizebox{8cm}{!}{\includegraphics{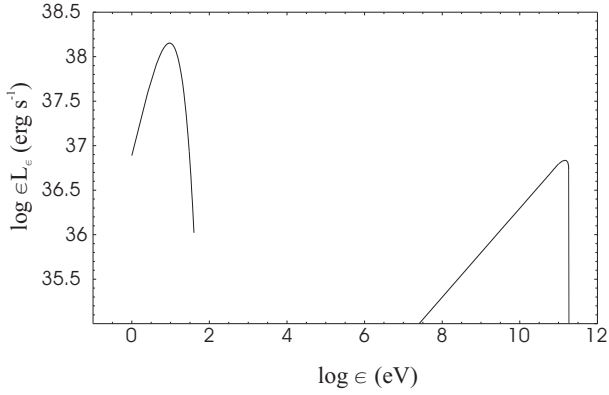}} \caption{\rm
Spectral high-energy distribution of the scattered photons for a
microquasar injecting a power-law spectrum of electrons in the
photon field of the high-mass stellar companion (an O7 star in
this example). A cut off at Lorentz factors $\gamma\sim10^3$ has
been assumed. The spectral energy distribution of the star is also
shown (left top corner).} \label{fig1}
\end{figure}

The luminosity is then obtained by integrating the scattering rate
(\ref{rate}) over the particle energy distribution (that can be
deduced using the density distribution given in Eq. (\ref{den})
and the volume of the blob), and multiplying by the observed
photon energy $\epsilon m_{e}c^{2}$ and the number density
$n_{p}=U/\epsilon_{0}m_{e}c^{2}$ (where \textit{U} is the energy
density of the photon field).
When $p$ is steeper than 1 and $\gamma_{\rm
min}=(\epsilon/4\epsilon_{0})^{1/2}\ll\gamma_{\rm
max}=\gamma_{2}D$, the luminosity reduces to:
\begin{eqnarray}
    L_{\epsilon}=\frac{dL}{d{\epsilon}d\Omega}\approx D^{3+p}\frac{kV\sigma_{T}c\:U2^{p-1}}{\pi
    \epsilon_{0}(1+p)(3+p)}\left(\frac{\epsilon}{\epsilon_{0}}\right)^{-(p-1)/2}
    \label{lumred}
\end{eqnarray}
In the Klein-Nishima regime numerical integrations are required
(see Georganopoulos at el. 2001b). The important point is that if
$\gamma_2\sim10^3$ then luminosities of $\sim 10^{36-37}$ erg
s$^{-1}$ can be obtained in the observer's frame at EGRET's energy
range (100 MeV- 20 GeV).

However, in the case of a single blob, the interaction time with
the photon field, and hence the duration of the gamma-ray flare,
will be strongly limited by the apparently superluminal speed of
the feature. For instance, in the case of an O7 star and a blob
with $\beta=0.98$ ($\Gamma=5$), the flux will decrease to $1/e$ of
its original value in $\sim 1$ s. Gamma-ray production by blobs in
microblazars are then transient phenomena difficult to detect on
Earth. This is not the case, however, of microblazars with
persistent jets.

For a continuous relativistic flow the amplification due to the
Doppler factor changes from $D^{3+p}$ to $D^{2+p}$ in Eq.
(\ref{lumred}) (Lind \& Blandford 1985). In this case the external
Compton scattering of stellar photons yields a stable source
inasmuch as the flux is not perturbed. We have calculated the
spectral energy distribution for a specific model with $p=2$, bulk
Lorentz factor $\Gamma=5$, viewing angle $\phi=10$ deg, photon
field of an O7 star (photon energies $\sim 10$ eV), and a
high-energy cut off of $\gamma_2=10^3>>\gamma_1$. The results are
shown in Figure 1, where we also show the energy distribution of
the stellar photons. We see that luminosities of $\sim 10^{36-37}$
erg $^{-1}$ can be obtained at $E>100$ MeV. Further details can be
introduced through a broken particle spectrum in orther to take
into account synchrotron cooling, but these refinements are not
necessary here.

It is interesting to notice that whereas the gamma-ray emission is
amplified by a factor $D^{2+p}=D^{3+2\alpha}$, the synchrotron jet
emission results amplified only by $D^{2+\alpha}$, where
$\alpha=(p-1)/2$ is the synchrotron spectral index (see Dermer
1995). Consequently, ``red'' microblazars, whose synchrotron
spectral energy distribution peaks at IR energies, will not be
ultra-luminous X-ray sources (ULXs) since only a small part of the
radiated power goes into the X-ray synchrotron tail of their
spectra and this emission, in addition, is not so strongly boosted
as the inverse Compton gamma-ray emission. Only in case of very
low values of $\gamma_2$ (say around $10^2$) we could get an X-ray
source ($L_X\sim 10^{33-34}$ erg s$^{-1}$) without a much more
energetic gamma-ray source.


The companion star in a high-mass microblazar system not only
provides a photon field for inverse Compton interactions, but also
a gravitational field that can exert a torque onto the accretion
disk around the compact object. The effect of this torque, in a
non-coplanar system, is to induce a Newtonian precession of the
disk. If the jets are coupled to the disk, as it is usually
thought, then the precession will be transmitted to them. This
situation, which should not be confused with the geodetic
precession (a purely General Relativity effect), has been
extensively studied in the case of SS433 (Katz 1980) and
extragalactic sources like 3C273 (e.g. Romero et al. 2000).

\begin{figure}
\resizebox{7cm}{!}{\includegraphics{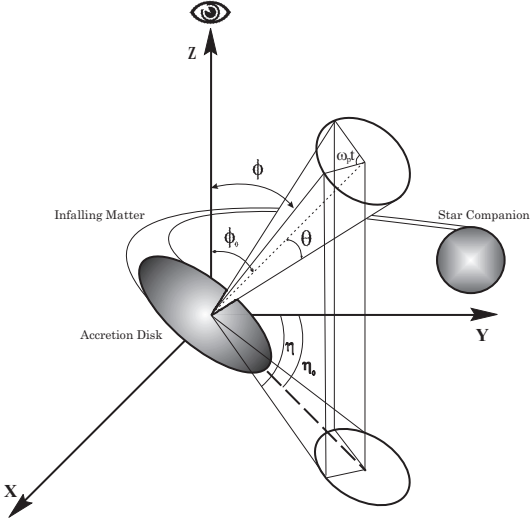}} \caption{\rm
Precessing jet model.} \label{fig2}
\end{figure}


\begin{figure}
\resizebox{6cm}{!}{\includegraphics{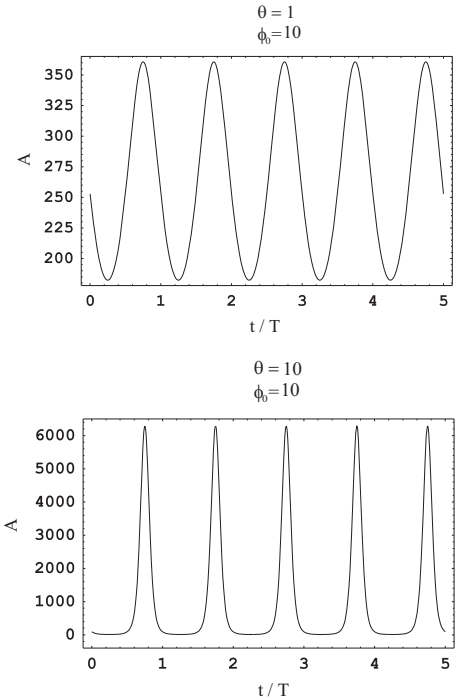}} \caption{\rm
Variation of the amplification factor for continuous jet emission
as a function of time in the precessing microblazar model (for two
different opening angles). Time units are normalized to the
precessing period.} \label{fig3}
\end{figure}

A sketch of the situation is presented in Figure 2. The disk
Keplerian angular velocity is $\omega_{\rm d}=(GM/r_{\rm
d}^3)^{1/2}$, where $M$ is the mass of the compact object and
$r_{\rm d}$ is the radius of the precessing part of the disk. The
orbital period is $T_{\rm m}$ and the orbital radius is given by
Kepler's law: $r_{m}^3=(G/4\pi^2)(m+M) T_{\rm m}^2 $, with $m$ the
mass of the star.

The angular velocity of the tidally induced precession can then be
approximated by $\omega_{\rm p}\approx -3Gm\cos\theta/4 r_{\rm
m}^3 \omega_{\rm d}$ (Katz 1980, Romero et al. 2000), where
$\theta$ is the half-opening angle of the precession cone. We can
now introduce a time-parametrization of the jet's viewing angle as
(see Abraham \& Romero 1999):
\begin{eqnarray}
    \phi(t)&=&\arcsin\left[x^{2}+y^{2}\right]^{1/2}\label{para}\\
    x&=&(\cos\theta\sin\phi_{0}+\sin\theta\cos\phi_{0}\sin\omega_{p}t)\cos\eta_{0}-\nonumber\\
    &&-\sin\theta\cos\omega_{p}t\sin\eta_{0}\nonumber \\
    y&=&(\cos\theta\sin\phi_{0}+\sin\theta\cos\phi_{0}\sin\omega_{p}t)\sin\eta_{0}+\nonumber\\
    &&+\sin\theta\cos\omega_{p}t\cos\eta_{0}\nonumber,
\end{eqnarray}
where all angles are defined in Fig. 2.

In Figure 3 we show the time evolution of the boosting
amplification factor of the gamma-ray emission for the case of a
continuous jet for two different sets of geometrical parameters
(viewing angle of 10 deg and precession half-opening angle of 1
and 10 deg). Time units are normalized to the precession period
$T$. We see that the flux density can change by a factor of
$\sim6\;10^3$ in a single period. A very weak, otherwise
undetected gamma-ray microblazar, can increase its flux due to the
precession and then enter within the sensitivity of an instrument
like EGRET, producing a variable unidentified gamma-ray source.
The duty cycle (i.e. the fraction of time in which the source is
highly variable) in this example is $\sim 0.2 T$, so the source
could appear in several EGRET viewing periods. Just to give a
feeling of the magnitudes involved, we mention that for a black
hole of 4 $M_{\sun}$, an O7Ia stellar companion, an orbital period
of 10 days, a half-opening angle of $10^{\circ}$ for the
precession, and an accretion disk of $\sim5\;10^{11}$ cm ($\approx
5\;10^{-2}$ a.u.), we get a precession period $T\sim100$ days. The
source could then be detectable $\sim70$ days per year. This is
consistent with some EGRET observations of highly variable sources
(e.g., the case discussed by Punsly et al. 2000).

\section{Discussion}

Taking into account that there exist more than $\sim130$ high-mass
X-ray binaries detected so far (e.g. Liu et al. 2000) and that
this number should be a small fraction of the total number of
these objects in the Galaxy, it is not unreasonable to expect the
existence of a few tens of microblazars at mid and low galactic
latitudes that could be responsible for the variable galactic
gamma-ray sources detected by EGRET. Moreover, the recently
discovered X-ray transient V4641 Sgr, which seems to harbor a
$\sim 9$ $M_{\sun}$-black hole and displays extreme superluminal
velocities, could be a microblazar (Orosz et al 2001). Since the
stellar companion is a late B-type star, external Compton
gamma-ray production is not expected to be very efficient in this
case, but its high-energy emission could fall within GLAST
sensitivity. Another interesting candidate, from the theoretical
point of view, is the high-mass X-ray binary LS I
$+61^{\circ}303$, which presents a one-sided jet at milliarsecond
scales and evidence for a precessing accretion disk (Massi et al.
2001). The intrinsic jet velocity, however, seems not to be very
high: $\sim0.4c$, but the source location is consistent with its
identification with a highly variable gamma-ray source 3EG
J0241+6103 (Tavani et al. 1998, Torres et al. 2001).

Not all microblazars should be confined to the galactic plane.
Recent direct measurements of proper velocities in both low-
(Mirabel et al. 2001) and high-mass (Rib\'o et al. 2002)
microquasars show that these objects can present very high
velocities and then could be ejected from the galactic plane. This
is also supported by the discovery of V4641 Sgr at $b\sim -4.8$
deg (estimated distance: $7.4\leq d \leq 12.3$ kpc). In the case
of microquasars with high-mass companions, which are young
objects, we could expect to find them up to distances $\sim100$ pc
(Rib\'o et al. 2002) or even more (e.g. V4641 Sgr) from the
galactic plane, so they could explain some of the mid-latitude
unidentified gamma-ray sources (Grenier 2001). A few low-mass
microquasars could have escaped to the galactic halo and
consequently might be related to some high-latitude unidentified
EGRET detections, if the jets are powerful enough.

Perhaps the best way to identified precessing gamma-ray
microblazars is through the detection of the electron-positron
annihilation feature in their spectra. This annihilation signature
should appear as a broad, blueshifted (by a factor $D$) line in
the spectrum at a few MeV, exactly within the energy range of IBIS
imager of the forthcoming INTEGRAL satellite. Due to the
precession of the jet, the Doppler factor will change periodically
with time, and hence the position of the annihilation peak should
oscillate in energy in the lab frame around a mean value (for a
detailed discussion of the phenomenon see Abraham et al. 2001).
Chandra X-ray observations of non-thermal radio sources within the
EGRET location error boxes (a complete list of these sources is
given by Torres et al. 2001) could help to find candidates to new
microquasars (through the detection of X-ray disk emission), and
then INTEGRAL exposures could be used in an attempt to find the
annihilation line. Such a detection would be a remarkable
discovery, since it would establish, at a same time, the matter
content of the jets in microquasars, and would help to clarify the
nature of some variable unidentified high-energy gamma-ray
sources.

\begin{acknowledgements}
We are very grateful to M. Georganopoulos for insightful
discussions and to the referee, M. Massi, for her valuable
comments. I.F.M. acknowledges support from grant PIP 0049/98 and
Fundaci\'on Antorchas. G.E.R. is supported by the research grants
PICT 03-04881 (ANPCT) and PIP 0438/98 (CONICET), as well as by
Fundaci\'on Antorchas. He is very grateful to staff of the Max
Planck Institut f\"ur Kernphysik at Heidelberg, where part of his
research for this project was carried out.

\end{acknowledgements}

{}

\end{document}